\documentclass[11pt,fleqn]{article}
\usepackage[utf8]{inputenc}
\usepackage{fullpage}

\usepackage{amsthm, amssymb}
\usepackage[dvipsnames]{xcolor}
\usepackage[hidelinks]{hyperref}
\usepackage{todonotes}
\usepackage{qcircuit}
\usepackage{tikz}
\usepackage{tikzsymbols}
\usetikzlibrary{calc}

\usepackage{titlesec}

\usepackage{pgfplots}
\pgfplotsset{compat=1.16}
\usepackage{svg}
\usepackage{mathtools}
\usepackage{titling}
\usepackage{textcomp}
\usepackage{mdframed}

\usepackage{titletoc}
\titlecontents{section}
  [1.5em]                                 %
  {}                             %
  {\contentslabel{1.5em}}                 %
  {}                                      %
  {\titlerule*[0.5pc]{.}\contentspage}    %

\usepackage{libertine}
\usepackage{libertinust1math}

\usepackage{graphicx, subfigure}
\usepackage{enumitem}
\usepackage{xparse}
\usepackage{physics}
\usepackage{color}
\usepackage{comment}
\usepackage{complexity}

\newcommand{\II}{\mathbb{I}}

\newcommand{\Exp}{\mathop{\mathbf{E}\hspace{0.13em}}}

\renewcommand{\Pr}{\mathop{\mathbf{Pr}\hspace{0.05em}}}

\newcommand{\eps}{\varepsilon}
\renewcommand{\epsilon}{\varepsilon}

\newcommand{\defeq}{\mathrel{\overset{\makebox[0pt]{\mbox{\normalfont\tiny\sffamily def}}}{=}}}

\usepackage{centernot}

\usepackage{complexity}

\renewcommand{\exp}{\mathsf{exp}}

\newcommand{\bits}{\{0,1\}}

\renewcommand{\tilde}{\widetilde}

\renewcommand{\eqref}[1]{\text{eq.~}(\ref{#1})}

\renewcommand{\exp}{\mathrm{exp}}

\usepackage{todonotes}

\newcounter{commentcount}

  {
  \begin{mdframed}
  [leftmargin=1em,rightmargin=1em,#1]\begingroup\small}%
  {\endgroup
  \end{mdframed}
  }

\usepackage{tikz}
\usetikzlibrary{shapes,backgrounds}

\usepackage{caption}
\allowdisplaybreaks 

\usepackage{tikz}
\usetikzlibrary{shapes,backgrounds}

\newtheorem{theorem}{Theorem}

\newtheorem{lemma}[theorem]{Lemma}

\newenvironment{xalign}[1][]{
    \subequations
    \label{#1}
    \align
}
{
    \endalign
    \endsubequations
}

\makeatletter
\let\oldendproof\endproof
\renewcommand{\endproof}{\oldendproof\par\noindent\ignorespaces}
\makeatother

\makeatletter
\let\oldendproof\endproof
\renewcommand{\endproof}{\oldendproof\par\noindent\ignorespaces}
\makeatother

\makeatletter
\renewcommand{\maketitle}{\bgroup\setlength{\parindent}{0pt}
\begin{flushleft}
  \LARGE {\@title} %
  \vspace{4.5mm}
  
  \normalsize \@author \linebreak
\end{flushleft}\egroup
}
\makeatother

\titleformat*{\section}{\large\bfseries}

\title{An alternate proof of the $\QMA$ lower bound for set-size estimation}

\author{Chinmay Nirkhe \\ \small University of Washington, \emph{Seattle, Wash.}}

\date{}

\begin{document}
\maketitle

\section{Introduction}
In the set-size estimation query task, a query algorithm decides whether an $N$-bit input has Hamming weight $\ge B$ or $\le A$ (with $\ge B$ as the yes instances and $\le A$ as the no instances). When $B = 2A$, this is sometimes called approximate counting. The query complexity for this problem with a quantum witness (i.e., $\QMA$ query bounds) was resolved in~\cite{aaronson_et_al:LIPIcs.CCC.2020.7} and proved using Laurent polynomials. Here we give a morally similar but distinct proof of the lower bound using the polynomial method.

\paragraph{Proof overview} A promising initial approach is to apply the OR degree bound to the symmetrized acceptance polynomial. However, it does not apply directly as soundness guarantees small values only on Hamming weights from $0$ to $A$, and this tail is too short to force the desired degree lower bound from the
jump at weight $B$. Instead, we observe that by applying the inclusion--exclusion principle, these same low-weight evaluations bound the coefficient of every monomial of degree at most $A$.
At $\theta_0=4B/N$, the total contribution of these monomials is small enough that deleting them leaves a univariate polynomial $h$ with $h(\theta_0)\ge 2^{-(A+3)}$. Since $h$ is divisible by $\theta^{A+1}$, the rescaled function $r(\theta)=(\theta_0/\theta)^{A+1}h(\theta)$ is itself a polynomial and satisfies $r(\theta_0)=h(\theta_0)$. The rescaling ensures that $r$ stays small on the entire interval $[64\theta_0,1]$, giving us the long tail needed to apply the OR bound.

\paragraph{Setup} The input is an $N$-bit string $x$ which is promised to have Hamming weight either $\ge B$ (yes instance) or $\le A$ (no instance) for integers $0 \le A < B \le N$.

\begin{theorem} Any $T$-query quantum algorithm with access to an $M$-qubit witness that correctly decides\footnote{An algorithm with access to a witness correctly decides a problem if for all yes instances, there exists a witness that causes the algorithm to accept with probability $\ge 2/3$ and if for all no instances, the maximum acceptance probability over all witnesses is $\le 1/3$.} all yes and no instances must satisfy either
\begin{equation}
    M \ge A + 1 \qquad \text{or} \qquad T = \Omega \qty( \frac{\sqrt{N/B}}{1 + \log \qty(\frac{B}{A+1})}).
\end{equation}
In particular for $B = 2A$, this bound is $M \ge A + 1$ or $T = \Omega(\sqrt{N/A})$.
\end{theorem}

\section{Proof}

\paragraph{Step 1: Amplifying success probabilities and removing the witness} Assume there exists a $T$-query and $M$-qubit witness algorithm deciding this problem with verifier circuit $V$. If $M < A$, pad the witness till it has length $M = A$ (as a longer witness can only help) and $B \le N/512$ (otherwise the theorem is trivial). Marriott-Watrous~\cite{Marriott2005} in-place amplification can be directly applied to the circuit, producing a circuit $V'$ of bounded error \begin{equation}\eps \defeq {\left(\frac{A+1}{2^{11}B}\right)^{A+1}}\end{equation} as opposed to that of $1/3$ for $V$. The circuit $V'$ requires $O((A+1)\log(2B/(A+1)))$ repetitions of $V$. Now, consider running the verifier $V'$ on the input $2^{-A} \II$; this is a technique also known as guessing the witness. The result is a witness-free quantum query algorithm whose acceptance probability (by~\cite{bealsetalpoly}) can be expressed as a degree $d$ multilinear polynomial $p$ where $d \le O((A+1)T\log(2B/(A+1)))$. The polynomial $p$ satisfies that
\begin{equation}
    (\textrm{a})~p(x) \in [0,1]~\forall~x \in \bits^N, \quad (\textrm{b})~p(x) \le \eps~\forall~ \abs{x} \le A, \quad (\textrm{c})~p(x) \ge (1 - \eps) 2^{-A} ~\forall~\abs{x} \ge B.
\end{equation}
Points (b) and (c) follow from the assumption of rejecting on all witnesses for no instances and accepting on at least one witness for yes instances, respectively.

\paragraph{Step 2: Truncating the polynomial to only high-degree terms}

Next, we use the fact that $p$ is small on low-Hamming-weight inputs (the soundness condition) to argue that the polynomial is well-approximated at $(\theta_0,\ldots,\theta_0)$, where $\theta_0 \defeq 4B/N$, by a polynomial $\tilde p$ which only consists of high-degree terms. To achieve this approximation, write out the multilinear polynomial $p$ as
\begin{equation}
    p(x) = \sum_{J \subseteq [N]} c_J x_J \quad \text{where~} x_J \defeq \prod_{i \in J} x_i.
\end{equation}
Then we claim that the coefficients $c_J$ can be extracted from polynomial evaluations using the inclusion-exclusion principle. Use $\mathbf{1}_U$ to denote the indicator vector of subset $U \subseteq [N]$. Then,
\begin{equation}
    c_J = \sum_{U {\color{RoyalPurple}\subseteq} J} (-1)^{\abs{J} - \abs{U}} p(\mathbf{1}_U).\label{eq:inclusion-exclusion}
\end{equation}
This inclusion-exclusion principle is proved easily via induction. Define then,
\begin{equation}
    \tilde p(x) = \sum_{J \subseteq [N] : \abs{J} > A} c_J x_J.
\end{equation}
Notice that when $\abs{J} \le A$,~\eqref{eq:inclusion-exclusion} implies that {\color{RoyalPurple}$\abs{c_J}$} $\le 2^{\abs{J}} \eps$. Since $\tilde p$ has deleted precisely these monomials, {at $x=(\theta_0,\ldots,\theta_0)$ we have}
\begin{xalign}[eq:bound-on-h]
    \abs{p(\theta_0,\ldots,\theta_0)-\tilde p(\theta_0,\ldots,\theta_0)}
    &\le \sum_{\abs{J}\le A}\abs{c_J}\theta_0^{\abs{J}}
    \le \eps\sum_{j=0}^A \binom Nj(2\theta_0)^j
     \le \eps\sum_{j=0}^A\frac{(8B)^j}{j!} \\
    &\le \eps\left(\frac{8eB}{A+1}\right)^{A+1}
     {=\left(\frac{e}{256}\right)^{A+1}} \le 64^{-(A+1)}.
\end{xalign}
For the second line, use $(8B/(A+1))^j\le(8B/(A+1))^{A+1}$ and $\sum_{j=0}^A (A+1)^j/j!\le e^{A+1}$.

\paragraph{Step 3: Succeeding on large Hamming weight inputs} In this step we exploit the completeness condition. Now for any choice of parameter $\theta \in [0,1]$, let $X_\theta$ be the distribution of $N$ i.i.d. Bernoulli variables with parameter $\theta$. By multilinearity,
\begin{equation}
    \Exp[p(X_\theta)] = p(\theta, \ldots, \theta) \in [0,1].
\end{equation}
If we define $h(\theta) \defeq \tilde{p}({\theta,\ldots,\theta})$, then $h$ is a univariate polynomial of degree at most $d$ with no terms of degree $\le A$. \\

\noindent Next, consider the behavior of $h$ at $\theta_0$, which equals $ 4 B /N$. By a Chernoff bound, the $\Pr[\abs{X_{\theta_0}} < B] \le \exp(-9B/8) < 1/2$ (a crude bound). Since the algorithm accepts every input of Hamming weight $\ge B$ with probability at least $(1-\eps)2^{-A} \ge 2^{-(A+1)}$, then $p(\theta_0, \ldots, \theta_0) \ge 2^{-(A+2)}$ and by~\eqref{eq:bound-on-h},
\begin{equation}
    h(\theta_0) = p(\theta_0, \ldots, \theta_0) + (\tilde{p}(\theta_0, \ldots, \theta_0) - p(\theta_0, \ldots, \theta_0)) \ge 2^{-(A+2)} - 64^{-(A+1)} \ge 2^{-(A+3)}.
\end{equation}

\paragraph{Step 4: The behavior of the univariate polynomial for large values}
The polynomial $h$ only consists of high-degree terms. We can therefore divide out a factor of $\theta^{A+1}$. Indeed, define
\begin{equation}
    r(\theta) \defeq \qty(\frac{\theta_0}{\theta})^{A+1} \cdot h(\theta).
\end{equation}
This is a univariate polynomial since $h(\theta)$ only consists of terms of degree $\ge A+1$. We next want to argue about the shape of $r$, observing that, by definition, $r(\theta_0) \ge 2^{-(A+3)}$, and that that for $\theta \in [64 \theta_0, 1]$, then
\begin{xalign}
    \abs{r(\theta)} &\le
      \qty(\frac{\theta_0}{\theta})^{A+1} \qty( \Exp\qty[p(X_\theta)]
      + 
      \Exp\qty[\abs{p(X_\theta)-\tilde p(X_\theta)}])\\
      &\le 64^{-(A+1)}+ \qty(\frac{\theta_0}{\theta})^{A+1} \sum_{\abs{J}\le A}\abs{c_J}\theta^{\abs{J}} \\
      &\le 64^{-(A+1)} + \sum_{\abs{J}\le A}\abs{c_J}\theta_0^{\abs{J}} \\
      &\le 2^{1-6(A+1)}
 \end{xalign}
where we applied~\eqref{eq:bound-on-h} for the last simplification.

\begin{figure}[htbp]
  \centering
  \begin{tikzpicture}[x=.94cm,y=.88cm,>=stealth,font=\small]
    \definecolor{rblue}{RGB}{30,77,112}
    \definecolor{rgreen}{RGB}{34,111,103}
    \fill[rgreen!13] (4.7,-.32) rectangle (10.1,.32);
    \draw[rgreen,densely dashed] (4.7,-.32) -- (10.1,-.32)
                                 (4.7,.32) -- (10.1,.32);
    \draw[rgreen!70] (4.7,-.32) -- (4.7,.32)
                     (10.1,-.32) -- (10.1,.32);
    \draw[->] (0,-.55) -- (0,2.9) node[above] {$r(\theta)$};
    \draw[->] (-.08,0) -- (10.8,0) node[right] {$\theta$};

    \draw[gray,densely dotted] (1.55,0) -- (1.55,2.05);
\draw[rblue,thick,<->,>=stealth]
  (.85,1.94) .. controls (1.08,2.01) and (1.32,2.05) .. (1.55,2.05)
  .. controls (2.65,2.05) and (3.35,.15) .. (4.7,.14)
  .. controls (5.7,.14) and (5.8,-.17) .. (6.8,-.13)
  .. controls (7.8,-.09) and (7.9,.17) .. (8.8,.12)
  .. controls (9.45,.09) and (9.75,-.06) .. (10.1,-.05);

\fill[rblue] (1.55,2.05) circle (2.2pt)
  node[above right=2pt] {$r(\theta_0)\ge 2^{-(A+3)}$};

    \node[rgreen] at (7.4,1) {$|r(\theta)|\le 2^{1-6(A+1)}$};
    \draw[->,rgreen] (7.4,.75) -- (7.4,.35);
    \node[below left] at (0,0) {$0$};
    \draw (1.55,.07) -- (1.55,-.07)
      node[below=2pt] {$\theta_0=\frac{4B}{N}$};
    \draw (4.7,-.32) -- (4.7,-.40) node[below] {$64\theta_0$};
    \draw (10.1,-.32) -- (10.1,-.40) node[below] {$1$};
  \end{tikzpicture}
  \caption{Cartoon of the polynomial $r(\theta)$.}
  \label{fig:set-size-r}
\end{figure}
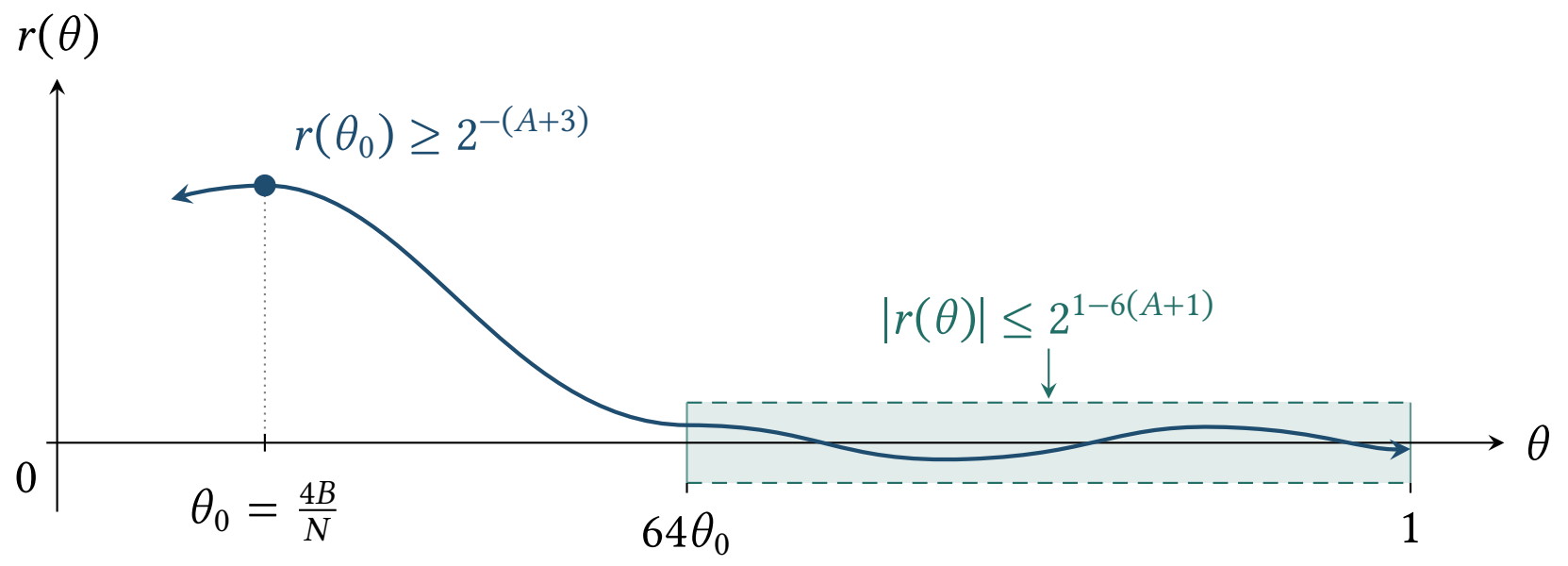

\paragraph{Step 5: Applying known polynomial degree lower bounds}

This implies that the polynomial $r(\theta)$ descends sharply and then stays small, as drawn in~Figure~\ref{fig:set-size-r}.
We can then apply well-known approximate degree lower bounds for the OR function (see~\cite{814607} for an example statement). We use the following version, which is the standard bound rescaled appropriately for this particular setting.

\begin{lemma}[Approximate degree bound for
$\mathrm{OR}$]\label{lem:continuous-or}
Let $A\ge 0$ be an integer and let $0<\theta_0\le 1/128$. Suppose a real
polynomial $r$ satisfies
\begin{equation}
    r(\theta_0)\ge 2^{-(A+3)}
    \qquad\text{and}\qquad
    |r(\theta)|\le 2^{1-6(A+1)}
       \quad\text{for every real }\theta\in[64\theta_0,1].
\end{equation}
Then $\deg r=\Omega\!\left((A+1)/\sqrt{\theta_0}\right)$.
\end{lemma}

\noindent The lower bound on $T$ follows as the degree of $r$ is at most $d = {O(T(A+1)\log(2B/(A+1)))}$.

\section{Sanity check: Where did we use the quantum witness?}

Observe that by the end of Step~1, we have removed the witness using the guess-the-witness technique. The remainder of the proof only studies the acceptance probability of a witness-free query algorithm. It is therefore worth asking, as a sanity check, why the same argument does not prove a query lower bound for $\mathsf{SBQP}$.

Consider the case $B=2A$, with $A\ge 1$. There is a one-query
$\mathsf{SBQP}$ algorithm: query a uniformly random bit of $x$ and accept if it is $1$. Its acceptance probability is at most $A/N$ on no instances and at least $2A/N$ on yes instances. This factor of two suffices for the definition of a $\mathsf{SBQP}$ algorithm. We can even see exactly where our proof would break
down for this particular algorithm. Its acceptance polynomial is given by the polynomial \(p(x)=\frac{1}{N}\sum_{i=1}^N x_i.\)
In our proof, Step~2 deletes every monomial of degree at most $A$, so it deletes this polynomial in its entirety. Equivalently, $h(\theta_0)=0$.

What prevents this from happening in our proof for $\QMA$ lower bounds? It is that the order of operations in
Step~1 matters. We first amplify the verifier \emph{while it still has the witness}, and only then run it on the maximally mixed state. This generates a polynomial of degree $\ge A$ which is not completely deleted.
Then steps~2 and~3 use both guarantees at the same point $\theta_0$. By
soundness, the total contribution of the deleted terms there has magnitude at most $64^{-(A+1)}$. By completeness and the choice of $\theta_0$, the original polynomial has value at least $2^{-(A+2)}$ there. The deleted terms therefore cannot account for the entire value of the polynomial.
This is precisely why $h(\theta_0)$ remains large enough to apply the degree bound.

An $\mathsf{SBQP}$ algorithm can also increase its multiplicative gap. For example, running the one-query algorithm $t$ times and accepting only if every run accepts increases the ratio between the two thresholds to $2^t$. But at Hamming weight $2A$ its acceptance probability is now $(2A/N)^t$. Each repetition also shrinks the quantity that Step~3 needs to survive truncation. In the $\QMA$ argument, amplifying before guessing is what preserves that quantity. Therefore, this provides a reproof of the oracle/query complexity separation between $\mathsf{SBQP}$ and $\QMA$.

\section{Comparison with the Laurent polynomial proof}

The result of this work is not new. Aaronson, Kothari, Kretschmer, and Thaler were the first to prove this statement~\cite{aaronson_et_al:LIPIcs.CCC.2020.7}. Our purpose here is to
give an alternate explanation. Indeed, the two proofs have much
in common. Both amplify the verifier and guess the witness, both average the acceptance probability over Bernoulli inputs, and both arrive at an ordinary univariate polynomial that is small on an interval and much
larger at a nearby point. From there, the same polynomial degree bound gives the query lower bound.

The difference lies in how the two proofs construct that univariate
polynomial. Aaronson \emph{et al.} consider the AND of two approximate-counting
instances. Averaging gives a bivariate polynomial $q$ in the two expected
Hamming weights, which they restrict to the hyperbola $\{(4At,4A/t)\}_t$.
At $t=1$ both expected weights are large. As $t$ increases, the second
expected weight becomes small, giving an interval on which the
restriction is small. The resulting Laurent polynomial is invariant
under $t\mapsto 1/t$ by symmetry between the two instances, so it is
an ordinary polynomial in $t+1/t$.

Our proof also uses reciprocal scaling. We use soundness on inputs of Hamming weight at most $A$ to
bound the low-degree coefficients by inclusion--exclusion. We can then
delete the terms of degree at most $A$ while preserving a large value
at $\theta_0$. The resulting $h(\theta)$ is divisible by $\theta^{A+1}$,
so the rescaling $r(\theta)$ gives an
ordinary polynomial in $\theta$. The same coefficient estimate shows
that $r$ is small on $[64\theta_0,1]$. This brings us back to the
univariate degree argument of Aaronson \emph{et al.}

\section*{Acknowledgments}
This work is supported by NSF CAREER Award IIS2541127. \\

\noindent Generative AI was used to simplify and check the presentation for correctness, but the main ideas were those of the author.

\bibliographystyle{alpha}
\bibliography{supplementary-files/refs}

\end{document}